\documentstyle[twocolumn,amstext,seceq,graphicx,amssymb]{jpsj}

\newcommand{\cP}{{\mathcal P}}

\newcommand{\cT}{{\cal T}}
\newcommand{\BEQ}[1]{\begin{equation}\label{#1}}
\newcommand{\EEQ}{\end{equation}}
\newcommand{\EQA}[1]{\begin{eqnarray}\label{#1}}
\newcommand{\PRL}[1]{Phys. Rev. Lett. {\bf #1}}
\newcommand{\PRB}[1]{Phys. Rev. B {\bf #1}}

\newcommand{\parc}{-J_H=10t=20J=20J_d}

\newcommand{\myfigwidth}{\linewidth}

\DeclareGraphicsExtensions{.eps} 
\graphicspath{{figures/}{./}}

\newcommand{\myfig}[3]{\begin{figure}[tb]
			  \begin{center} 
			\includegraphics*[width=\myfigwidth]{#1}
			\caption{#3}
			\label{#2}
	                \end{center}
		\end{figure}}

\newcommand{\figasgap}{
	\myfig{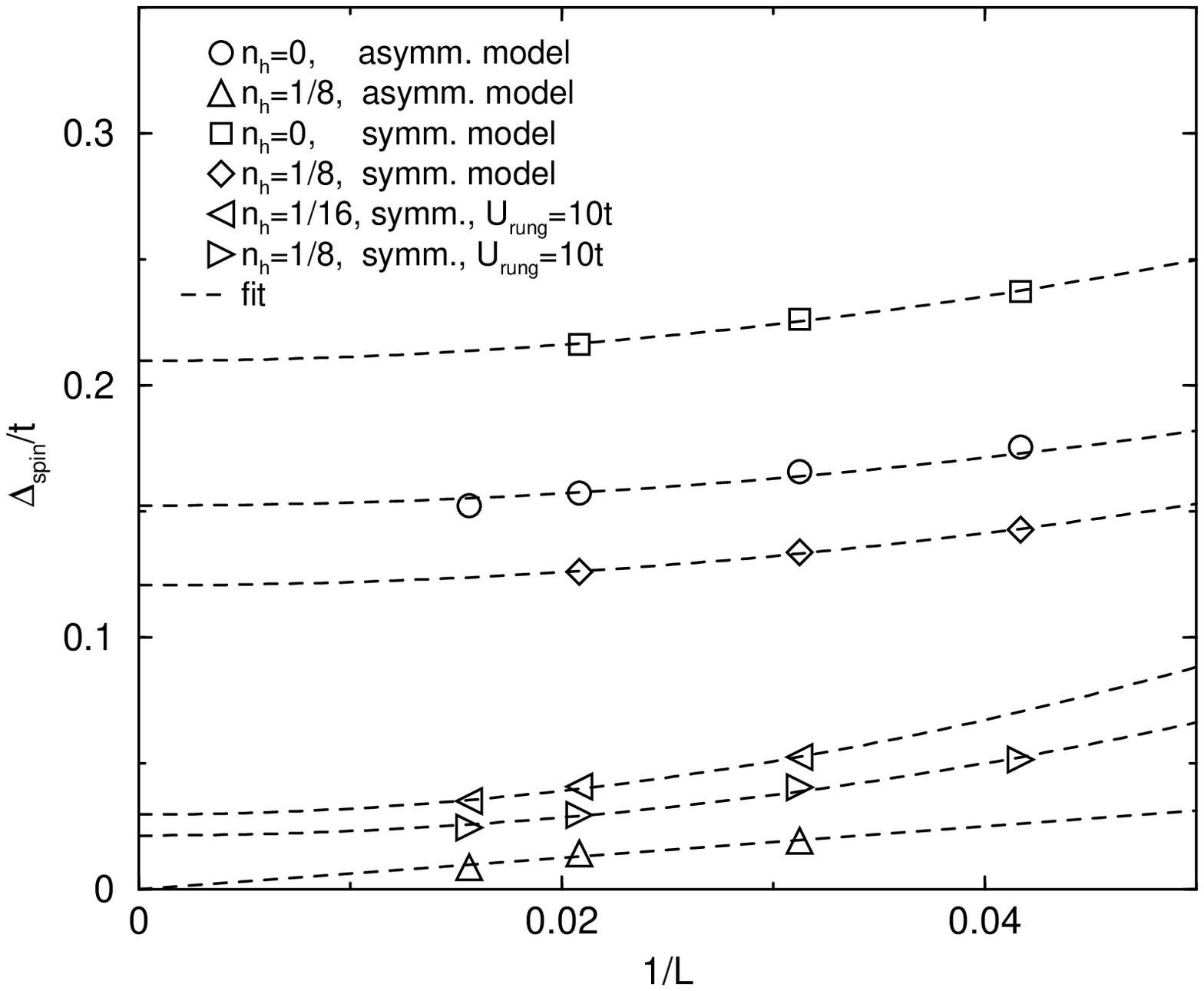}{FigGap}{Finite-size scaling of the spin gap
	$\Delta_s$ for the different models with $\parc$, and $U=10t$
	for the model with Coulomb repulsion, and different hole
	densities of $n_h=0$, $n_h=1/8$, and $n_h=1/16$.}
}

\newcommand{\figchi}{
	\myfig{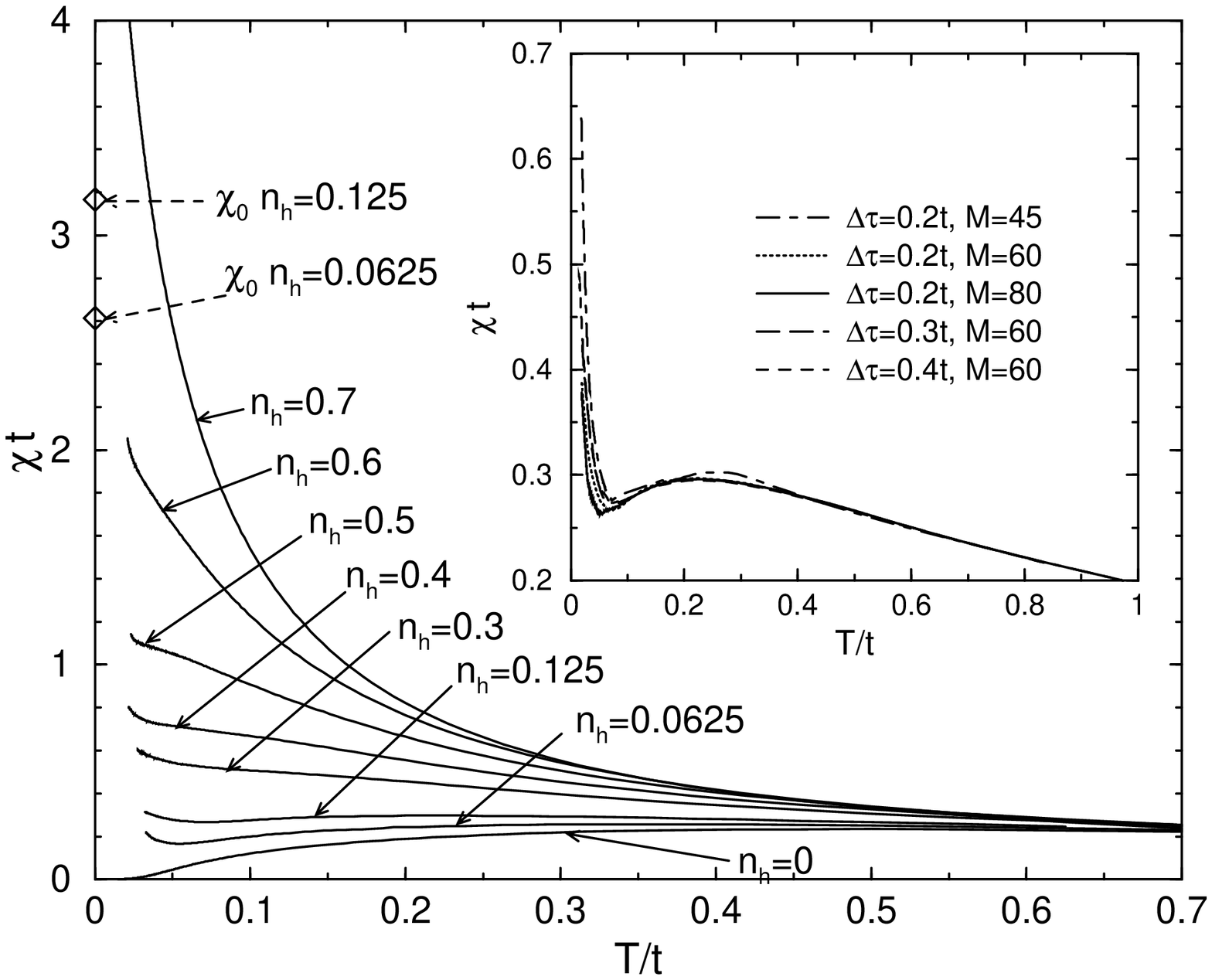}{FigChi}{Magnetic susceptibility $\chi$ of the
	asymmetric model for $\parc$. The $T=0$ values are obtained
	from the spin velocities obtained by a finite-size scaling
	analysis. In the inset we show the convergency of the 
	TDMRG method for different sizes of the Trotter time
	steps $\Delta\tau$ and different numbers of states kept $M$ 
	for $n_h=0.125$.}
}

\newcommand{\figsp}{
	\myfig{fig3}{FigSpSp}{Spin-spin correlation functions
	$S_i^zS_j^z$ at low hole doping $n_h=0.125$ for the different
	models calculated by the DMRG method on $2 \times 128$ site
	systems and $\parc$, and $U=-J_H$ for the symmetric model with
	Coulomb repulsion. For comparison we also show the undoped
	symmetric system. } }

\newcommand{\figstring}{
	\myfig{fig4}{FigString}{String correlation function
	$g(x_0,x)=\langle (\sum_{i=1,2} S_{x_0,i}^z)$\newline$\left(
	\prod_{k=x_0+1, j=1,2}^{x-1} e^{i \pi S_{k,j}^z} \right)
	(\sum_{l=1,2} S_{x,l}^z)\rangle $ at low hole doping for the
	different models with $\parc$ and $U=-J_H$ for the symmetric
	model with Coulomb repulsion obtained by DMRG calculations on
	systems with $2\times 128$ sites.}  }

\newcommand{\figch}{
	\myfig{fig5}{FigCh}{
	Friedel oscillations in the charge 
	density distribution due to the open boundary conditions for the
	different models with $\parc$ at low hole doping $n_h=0.125$. 
	In the inset we show the corresponding Fourier transform.}
}

\newcommand{\figtriplpair}{
	\myfig{fig7}{FigTripl}{Triplet pairing correlations
	   $P_{i}(f)P_{i+x}^{\dagger}(f')$ with $P_{i}^\dagger(f)=
	   \frac{1}{\sqrt{2}}(c_{i,\uparrow}^\dagger c_{i+f,\downarrow}^\dagger
	   +c_{i,\downarrow}^\dagger c_{i+f,\uparrow}^\dagger)$ 
	for (a) symmetric model, (b) symmetric model with
	Coulomb repulsion $U=10t$, and (c) asymmetric model at
	$n_h=1/8$ and $\parc$.
	}
}

\newcommand{\figpair}{
	\myfig{fig6}{FigPair}{Singlet pairing correlations
	   $P_{i}(f)P_{i+x}^{\dagger}(f')$ with $P_{i}^\dagger(f)=
	   \frac{1}{\sqrt{2}}(c_{i,\uparrow}^\dagger c_{i+f,\downarrow}^\dagger
	   -c_{i,\downarrow}^\dagger c_{i+f,\uparrow}^\dagger)$ 
	for (a) symmetric model, (b) symmetric model with
	Coulomb repulsion $U=10t$, and (c) asymmetric model at
	$n_h=1/8$ and $\parc$.
	}
}

\newcommand{\figspsm}{
	\myfig{fig8}{FigSpSm}{Spin-spin correlations in the $x-y$
	plane for the asymmetric model and symmetric model 
	for $S^z_{\text{tot}}=0$ with $J_H=-10t$, $L=32$, $n_h=0.125$
	and various values of $J,J_d$, calculated by the DMRG method.
	} 
}

\newcommand{\figFMPair}{
	\myfig{fig9}{FigFMPair}{Pairing corrleation functions
	$P_{i}(f)P_{i+x}^{\dagger}(f')$ with $P_{i}^\dagger(f)=
	\frac{1}{\sqrt{2}}(c_{i,\uparrow}^\dagger
	c_{i+f,\downarrow}^\dagger \mp c_{i,\downarrow}^\dagger
	c_{i+f,\uparrow}^\dagger)$in the FM phase for (a) symmetric
	model,and (b) asymmetric model at $n_h=1/8$,$J_H=-10t$
	and $J=J_d=0.01t$.  } }

\def\runtitle{Doped two orbital chains with strong Hund's rule couplings}
\def\runauthor{Beat {\sc Ammon} and Masatoshi {\sc Imada}}

\kword
{
spin-$1$ systems, Mott-Hubbard transition, superconductivity, 
triplet superconductivity, ferromagnetism, string order
}
\title{Doped two orbital chains with strong Hund's rule couplings -
	ferromagnetism, spin gap, singlet and triplet pairings}
\author
{ 
Beat {\sc Ammon} \footnote{E-mail address: ammon@issp.u-tokyo.ac.jp}
and Masatoshi {\sc Imada} \footnote{E-mail address: imada@issp.u-tokyo.ac.jp}
}

\inst
{
ISSP, University of Tokyo, 7-22-1 Roppongi, Minato-ku, Tokyo 106-8666, Japan
}

\recdate
{
\today
}

\abst{
Different models for doping of two-orbital chains with mobile $S=1/2$
fermions and strong, ferromagnetic (FM) Hund's rule couplings
stabilizing the $S=1$ spins are investigated by density matrix
renormalization group (DMRG) methods. The competition between
antiferromagnetic (AF) and FM order leads to a rich phase diagram with
a narrow FM region for weak AF couplings and strongly enhanced triplet
pairing correlations. Without a level difference between the orbitals,
the spin gap persists upon doping, whereas gapless spin excitations
are generated by interactions among itinerant polarons in the presence
of a level difference. In the charge sector we find dominant singlet
pairing correlations without a level difference, whereas upon the
inclusion of a Coulomb repulsion between the orbitals or with a
level difference, charge density wave (CDW) correlations decay
slowest. The string correlation functions remain finite upon doping
for all models.
}

\begin{document}
\maketitle

\section{Introduction}
A key problem in the research on the Mott-Hubbard transition consisits
of the doping of a spin liquid groundstate with mobile holes. The
antiferromagnetic (AF) $S=1$ Heisenberg (HB) spin chain is an example
of such a spin liquid with a finite spin gap \cite{Haldane} and a
hidden topological order, the string correlation
function \cite{StringOrder}. Doping of $S=1$ HB chains poses a number
of interesting theoretical questions caused by the competition between
ferromagnetic (FM) order induced by the double exchange mechanism
\cite{DblExch} and AF order which can result in completely
different magnetic properties. Another relevant question is whether
the spin gap is destroyed immediately upon doping or whether it
persists, and the competition also determines which correlation
function characterized by the single correlation exponent $K_\rho$
dominates in the thermodynamic limit of infinite system size.

Only recently has it become possible to experimentally study mobile
holes doped in a $S=1$ chain in the system $\rm Y_{2-x}Ca_xBaNiO_5$
\cite{DiTusa}. The two active $\rm Ni^+$ orbitals are the
$3d_{3z^2-r^2}$ and the almost localized $3d_{x^2-y^2}$ orbital. 
Strong Hund's rule couplings $J_H$ between these two orbitals and very
weak inter-chain couplings make this system an almost ideal $S=1$ HB
chain with a spin gap of $\Delta_s \approx 100K$ in the undoped case
\cite{YBaNiO}. By replacing off-chain $\rm Y^{3+}$ with $\rm Ca^{2+}$,
mobile holes can be introduced into the system and $\mu$SR data show
that these carriers indeed have spin $S=1/2$ \cite{Kojima}. The
most interesting experimental features upon doping consist of a
reduction of the resistivity $\rho_{\text{dc}}$ by several orders of
magnitude and that the temperature dependence of $\rho_{\text{dc}}$ no
longer can be described by thermal activation over a barrier
\cite{DiTusa}. Further new states with spin $S$ between $1$ and $3/2$ per
impurity appear below the Haldane gap \cite{DiTusa}.

First theoretical studies on doped $S=1$ chains concentrated on
localized $S=1/2$ impurities \cite{Sorenson}. For sufficiently weak
couplings, they give rise to bound $S=1/2$ states below the
Haldane-gap \cite{Kaburagi}, similar to the chain-end excitations
\cite{Kennedy}. However, it soon turned out that mobile carriers are
needed for an accurate description of this system and the case of weak
hopping \cite{Penc} has been studied first. A more realistic effective
Hamiltonian for fully mobile holes has been derived in \cite{Dagotto}
for one- and two-band models with infinitely strong Hund's rule
couplings. A phase diagram which shows a FM region and phase
separation for large values of the coupling $J$ between the $S=1$
spins has been obtained by exact diagonalization and (DMRG) studies of
small systems \cite{Riera}. This study also suggests a possible
hole-bound region between the FM and phase separated ones. Based on
DMRG calculations for very large systems, we briefly have reported in
a previous Letter on a new hierarchy of energy scales for a model with
finite Hund's rule couplings including fully mobile holes and
localized electrons in the lower band \cite{Spin1}. Ferromagnetic
polarons formed by the double exchange mechanism are found to weakly
interact in the lower energy scale in the background of the gapped
spin liquid formed at the higher energy. Moreover, doping-induced
magnetization steps have been discovered for an exactly solvable model
which includes biquadratic exchange \cite{Frahm}. The biquadratic
terms are needed for the solution by the algebraic Bethe Ansatz.

While all of the above mentioned studies are restricted to models with
mobile electrons in the upper band only, the very different behavior
of systems with and without a level difference between the two
orbitals has first been investigated by weak-coupling theories
\cite{Fujimoto,Nagaosa}. These studies show that the spin gap remains
finite if there is no level difference between the orbitals, whereas
the spin gap is destroyed immediately if the particle density between
the two orbitals differs. We have recently carried out the first study
of this problem in the strong coupling regime by numerical methods and
discovered dominant superconducting pairing correlations for systems
without a level difference in contrast to dominant charge-density wave
(CDW) correlations for systems with a level difference
\cite{LevelDiff}. This surprising result which exhibits completely
different phases depending on the level difference emphasizes the
importance of a detailed study of realistic models in the strong
coupling regime. Unfortunately this is also the most difficult regime
to investigate, and it is not clear whether additional terms need to
be included in a realistic model.

In this paper we present a detailed analysis of a realistic model in
the strong coupling regime with strong but finite Hund's rule
couplings $J_F$ and fully mobile holes. Based on large scale numerical
calculations by both the thermal DMRG (TDMRG) \cite{TDMRG} and ground
state DMRG method \cite{White,WhiteFullDMRG}, we will to cover what
has been left out due to lack of space in our two previous
publications. In addition, we include a model with a Coulomb repulsion
and no level difference between the electrons in the two orbitals on
the same site.

We start with the discussion of a generic model for $\rm
Y_{2-x}Ca_xBaNiO_5$ which includes a level difference between the two
orbitals, and we will call this the asymmetric model in the
following. In this case, the electrons in the lower $3d_{x^2-y^2}$
orbital are almost localized, and all holes are doped into the higher
lying $3d_{3z^2-r^2}$ orbital. We describe these mobile holes in the
upper band by the following Hamiltonian:
\begin{equation}\label{Hkin}
H_{\text{kin}}^{(i)} = -t \sum_{j,\sigma} \cP \left(
        c^{\dagger}_{j,i,\sigma} c_{j+1,i,\sigma}
        +H.c.\right) \cP ,
\end{equation}
where the projection operator $\cP$ prohibits doubly occupied sites,
$c_{j,i,\sigma}^\dagger$ is the particle creation operator on site $j$
in the orbital $i$ with spin $\sigma$, and the index $i=1$ denotes the
upper orbital and $i=2$ the lower orbital. Due to virtual hopping
processes there exist AF couplings $J>0$ between nearest-neighbors in
the upper band, and we include further AF couplings $J_d>0$ between the
electrons in different orbitals on neighboring sites as follows:
\begin{eqnarray} \label{Haf}
H_{\text{af}}^{(i)}& =& J \sum_{j} \left(\vec{S}_{j,i} \vec{S}_{j+1,i}
		-{1\over4}n_{j,i}n_{j+1,i}\right) \\
H_{\text{diag}}& =& J_d \sum_{j} \left(\vec{S}_{j,1} \vec{S}_{j+1,2}
	+\vec{S}_{j,2}\vec{S}_{j+1,1} -{1\over2}n_{j,1}n_{j+1,2}\right), \nonumber \\
\label{Hdiag}
\end{eqnarray}
where the indices are the same as before, $n_{j,i}=
c^{\dagger}_{j,i,\uparrow}c_{j,i,\uparrow}
+c^{\dagger}_{j,i,\downarrow}c_{j,i,\downarrow}$, and the rest of the
notation is standard. The strong Hund's rule coupling $J_H<0$ between
the electrons in different orbitals on the same site reads:
\begin{equation}\label{HHund}
H_{\text{FM}} = J_H \sum_{j} \vec{S}_{j,1} \vec{S}_{j,2},
\end{equation}
and is responsible for the formation of the $S=1$ spins. Finally the
Hamiltonian for a model with a level difference is expressed as:
\begin{equation}\label{Has}
H_{\text{as}} = H_{\text{FM}} + H_{\text{diag}} +
		H_{\text{af}}^{(1)} + H_{\text{kin}}^{(1)}.
\end{equation}
This model is easy to understand in two limiting cases: first at half
filling, where it can be mapped to a Haldane $S=1$ chain with a spin
liquid ground state and a finite spin gap. The second case is away
from half-filling at $J=J_d=0$, where all the spins are
ferromagnetically aligned due to the double-exchange mechanism in
order to gain kinetic energy. Hence, for finite values of $J,J_d$
there is a competition between FM order induced by the double exchange
mechanism and the spin liquid ground state. For $|J_H| \gg J, J_d$
each hole is surrounded by a small FM cloud created by the double
exchange mechanism, and we will call this a polaron in the
following. An intriguing question on the doped system for small $J$
and $J_d$ is the possiblity of triplet pairing. In this paper we show
a strong enhancement of the triplet pairing corrlelations.  However,
for $J,J_d>0.1t$, ferromagnetic polaron becomes only a weak
perturbation of the underlying spin liquid which remains intact. Among
these polarons $2k_F$ and $4k_F$ charge density wave (CDW) order is
then stabilized. A hierarchy of energy scales is established in the
spin sector with the smaller scale given by the gapless, lowest-lying
interactions among the polarons; and the second, larger energy-scale
consists of the spin liquid background.
	
On the other hand, we can think of a model with a very small or no
level difference at all. This situation is more likely in a
ladder-system, and we will refer to this model as the symmetric model
in the following. From the symmetric model we include hopping and AF
couplings $J>0$ in both orbitals, and the Hamiltonian is given by:
\begin{equation}\label{Hsym}
H_{\text{sym}} = H_{\text{FM}} + H_{\text{diag}} + \sum_i \left(
	H_{\text{af}}^{(i)} + H_{\text{kin}}^{(i)} \right) .
\end{equation}
In this model the double exchange mechanism is expected to be less
effective, since hole pairs on the same rung can gain the large Hund's
rule coupling $J_H$. However, in many realistic situations there
might be an additional Coulomb repulsion between the electrons on the
same rung, and we will include this by the following additional term:
\begin{equation}\label{Hrep}
	H_{\text{rep}} = U \sum_{j} n_{j,1} n_{j,2}.
\end{equation}
This Coulomb repulsion will reestablish the competition between the
double-exchange mechanism and the spin liquid ground state, while
preserving the same particle density in both orbitals.

For all the models we restrict our investigations to the case of 
strong Hund's rule couplings $|J_H| \gg |J|,|J_d|,|t|$, and if 
not otherwise mentioned we always set $\parc$, and $U=|J_H|$ for
the model with a Coulomb repulsion between the electrons on the
same site.

The rest of this paper is organized as follows: we start with a
discussion of the numerical methods first. In the next section we
investigate various physical quantities in detail, and we begin with
the discussion of the spin gap. There we see that the spin gap is
destroyed immediately upon doping for the asymmetric model, while it
remains finite for the symmetric models. Consistent with these
findings we obtain finite values of the magnetic susceptibility $\chi$
for $T\rightarrow 0$ for the asymmetric model. We continue with the
study of several correlation functions obtained by the DMRG method,
and from the correlation exponent $K_\rho$ obtained by fitting to the
asymptotic form of a Tomonaga-Luttinger liquid in the gapless and a
Luther-Emery liquid in the gapful case, we find that superconducting
pairing correlations dominate in the thermodynamic limit of infinite
system size for the symmetric model without Coulomb repulsion, while
CDW correlations dominate in the two other cases. Finally, we
investigate ferromagnetism and triplet pairing correlations for small
values of the AF couplings $J,J_d$, and the last section of this
report is devoted to the conclusion.

\section{Numerical Methods}
In order to take the strong correlation effects into account properly,
unbiased and approximation free methods are required, and large
system-sizes are needed to avoid finite-size effects. For the
investigation of the thermodynamic properties, we use the novel TDMRG
me\-thod \cite{TDMRG}. This method is based on the combination of the
transfer matrix method \cite{TM} and the DMRG method. The TDMRG method
allows us to reach temperatures as low as $T=0.02t$ in the
thermodynamic limit of infinite system size; and the only two sources
of errors stem from the truncation errors of the DMRG procedure and
the finite size of the Trotter time steps.

In order to apply the transfer matrix method we first map a
one-dimensional quantum system to a two-di\-mensional classical system
by the Trotter-Suzuki decomposition \cite{trotter_suzuki}. For this
purpose we decompose the Hamiltonian into two parts
$H=H_{\text{odd}}+H_{\text{even}}$ where each
$H_{\text{odd}(\text{even})}$ is a sum of commuting terms
$H_{\text{odd}}=$
\newline
$\sum_{i=1}^{L/2} h_{2i-1,2i}$ and
$H_{\text{even}}=\sum_{i=1}^{L/2} h_{2i,2i+1}$. We then write the
partition function of a system with $L$ sites as 
\begin{equation}
	Z_{L} = \text{tr} e^{-\beta H} 
	= \lim_{N\rightarrow \infty} Z_{N,L}
\end{equation} 
with $Z_{N,L}=\text{tr} \left( e^{-\beta H_{\text{odd}}/N}
e^{-\beta H_{\text{even}}/N} \right)^N$ and $\beta=1/T$. With the
definition of the virtual transfer-matrix $\cT_i=\left( e^{-\beta
h_{2i-1,2i}/N} e^{-\beta h_{2i,2i+1}/N} \right)^N$ we obtain
$Z_{N,L}=$
\newline
$lim_{N\rightarrow \infty} \text{tr} \prod_{i=1,L}
\cT_i$. By applying the Trotter-Suzuki interchangeability theorem
\cite{trotter_suzuki} we can exchange the limit of the system size
with the limit of the Trotter number and obtain the free energy
density $f$ in the thermodynamic limit of infinite system size
\begin{eqnarray}\label{TransferMat}
f &=& lim_{N\rightarrow \infty} lim_{L\rightarrow \infty}
	\left(-\frac{1}{\beta L}\right)\ln \text{tr} 
	\prod_{i=1,L} \cT_i \nonumber \\
& =& lim_{N \rightarrow \infty}  \left(- \frac{1}{\beta}\right) 
	\ln \lambda_{\text{max}},
\end{eqnarray}
where $\lambda_{\text{max}}$ is the maximum eigenvalue of $\cT_i$. We
can then calculate the maximum eigenvalue $\lambda_{\text{max}}$
numerically and obtain physical quantities from derivatives of $f$.
In the actual calculations, several thermodynamic quantities such as
the internal energy $u$ or the magnetization $m$ are obtained directly
from the maximal eigenvector of $\cT_i$ \cite{TM,troyer_tm}. For a
fixed value of the Trotter time steps $\Delta\tau=\beta_0/N_0$, we
then lower the temperature $T$ by inserting additional time steps
along the imaginary time direction between the system and environment
parts of the transfer matrix $\cT_i$. At the same time the dimension
of the transfer matrix $\cT_i$ is kept tractable by applying the DMRG
method and truncating the states with the smallest weight in the
density matrix. The dimension of the Hilbert space can further be
reduced by using symmetries such as the spin conservation symmetry and
performing the calculations in the subspace of zero winding number.

In our calculations we usually keep $M=80$ states for each of the
system and environment blocks of the DMRG procedure, and estimate the
truncation errors by comparing to the results with $M=40$ and $60$
states. For $M=60$ states, the truncation error $\varepsilon$ is
always smaller than $10^{-3}$. Another source of errors stems from the
finite size of the Trotter time steps $\Delta \tau$. Basically one can
extrapolate to $\Delta \tau\rightarrow 0$ by performing calculations
for several different values of $\Delta \tau$. However, for doped
systems, the TDMRG calculations need to be performed in the grand
canonical ensemble, and the chemical potential $\mu$ is fixed for each
DMRG iteration. Thus we need to extrapolate to constant particle
density $n$ afterwards, and for a large number of Trotter time steps
$N$, the errors introduced by this extrapolation to constant particle
density are in general larger than the errors caused by the finite
size of the Trotter time steps. If not otherwise mentioned, we have
therefore performed the calculations for fixed $\Delta \tau =0.2t$ and
neglected the errors caused by finite Trotter time steps. We have also
used a re-biorthogonalization algorithm if we encounter numerical
instabilities in the TDMRG procedure \cite{BiOTDMRG}.

Complementary to the TDMRG method, we have used the ground state DMRG
method to investigate the correlation functions and ground-state
energies. The main advantage of this method is that it allows very
accurate calculations of these quantities for large $1D$ systems. We
have investigated systems of up to $2\times256$ sites and kept up to
$M=1400$ states for the system and environment block each. The
truncation errors encountered in our calculations are usually smaller
than $\varepsilon < 10^{-8}$. All calculations have been performed on
Alpha workstations at the ISSP.

%
\section{Results}
\subsection{Spin gap and magnetic properties}
We start our investigations with the spin gap properties, and
determine whether it remains finite upon doping for the different
models. In the undoped case, both models can be mapped to the Haldane
$S=1$ chain for $J_H \gg J,J_d$. The effective coupling
$J_{\text{eff}}$ between the $S=1$ spins is
$J_{\text{eff}}^{\text{as}}=3J/4$ for the asymmetric and
$J_{\text{eff}}^{\text{sym}}=J$ for the symmetric model by second
order perturbation theory, thus we expect a spin gap of
$\Delta_s=0.41050(1) J_{\text{eff}}$ in both cases.

By finite-size scaling we have numerically determined the spin gap
from $\Delta_s=\lim_{T\rightarrow 0} \Delta_{s}(L;N=Ln)$, where
$\Delta_{s}(L;N=Ln)=\Delta_{s}(L;N)=E_0(L;N;S^z=1)-E_0(L;N;S^z=0)$
with $E_0(L;N;S^z)$ denoting the ground-state energy of the system
with $N$ particles on $L$ sites and total spin component along the
$z$-direction $S^z$. The result of the finite-size scaling analysis by
DMRG calculations (see Fig.~\ref{FigGap}) is in excellent agreement
with the above mapping, and we obtain $\Delta_s=0.41(1)
J_{\text{eff}}^{\text{as}}\approx 0.154(5)t$ in the asymmetric case
and $\Delta_s=0.41(1) J_{\text{eff}}^{\text{sym}}\approx 0.205(5)t$
for the symmetric model.
\figasgap
In agreement with refs.~\cite{Fujimoto,Nagaosa} we also find
completely different behavior for both models upon doping, as is
evident from Fig.~\ref{FigGap} and we have listed the results in
Table~\ref{TabGap}. For the asymmetric model, the spin gap is
completely destroyed already for hole densities of $n_h=0.0625$ and
$n_h=0.125$, where $n_h$ denotes the hole density in the conduction
band. On the other hand, the gap is reduced, but remains finite for
the symmetric model for the same hole densities $n_h=0.0625$ and
$n_h=0.125$, where we set $n_h=1-n$ with $n$ denoting the number of
electrons per orbital for the symmetric models. The reduction of the
spin gap is stronger in the symmetric case including a Coulomb
repulsion $U=J_H$, but appears to remain finite. However, great care
needs to be taken with the boundary conditions of the DMRG
calculations in order to avoid the degenerate $S=1/2$ spin excitations
at the end of the open chain, which would result in spin gaps
exponentially small with the system-size $L$ \cite{White}. By
enumerating the sites of the open chain from $i=1$ to $L$, we have
used the boundary condition which enforces a $S=1/2$ spin at site
$i=1$ and $S=1$ at site $i=2$, and similarly at the other end of the
chain. If we do not enforce a $S=1$ spin next to the $S=1/2$ spins we
find that holes are trapped at the boundary and the finite-size
scaling becomes unreliable or impossible for smaller system sizes
because of chain-end excitations.
\begin{fulltable}
\begin{tabular}{l|cc|cc|cccc}
$n_h$ & $\Delta^{\text{sym}}_s$ & $\xi^{\text{sym}}$ &
	 $\Delta^{\text{sym+re}}_s$ & $\xi^{\text{sym+re}}$ &
	 $\Delta^{\text{as}}_s$ & $\xi^{\text{as}}$ & 
		$v^{\text{as}}_{\sigma}$ & $\chi_0$ 
\\ \hline
0      &  $0.205(5) t$ &  $6.05(5)$ &   $0.205(5) t$ &  $6.05(5)$ 
		&  $0.1504(5) t$ & 	$6.05(5)$  & - & 0 \\
0.0625 &  $0.148(5) t$ & $6.1(2)$ & $0.03(1) t$ &  $15.4(5)$
		&  0  & $7.9(2)$ &  0.244 & 2.61\\
0.125  &  $0.120(5) t$ &  $6.7(2)$ & $0.02(1) t$ & $20.5(5)$ &
		0 & $11.2(2)$ & 0.201 & 3.17
\end{tabular}
\caption{Spin gap obtained by finite-size scaling analysis for the 
	symmetric model (left), symmetric model with Coulomb repulsion
	$U=10t$ (middle), and the asymmetric model (right) for
	$\parc$. Also listed is the spin-spin correlation length $\xi$,
	and for the gapless asymmetric model the spin velocity
	$v_{\sigma}$ and magnetic susceptibility $\chi_0$ at $T=0$. 
	The definition of $\xi$ is not simple for the cases with 
	$\Delta_s=0$, see text for details.}
 \label{TabGap}
\end{fulltable}

From the finite-size scaling we can also determine the spin velocity
and magnetic susceptibility $\chi$ at $T=0$ of the gapless asymmetric
model by noting that the smallest momentum in the finite open chain is
$k_{\text{min}}=\pi/L$. The spin velocity $v_\sigma$ can then be
obtained from $\Delta_s(L)=v_{\sigma}k_{\text{min}}$, and the magnetic
susceptibility is given by $\chi_0=\frac{2K_\sigma}{\pi v_\sigma}$,
where $K_\sigma=1$ due to the SU(2) symmetry. We see from 
Table~\ref{TabGap} that the magnetic susceptibility $\chi$ is rather large
already for small values of hole doping.

Next we compare these findings for the magnetic susceptibility with the
TDMRG results shown in Fig.~\ref{FigChi} for the asymmetric model for
$\parc$, and we refer to Fig.~1 of ref.~\citen{LevelDiff} for
the symmetric model. For the symmetric model we have set $-J_H
\rightarrow \infty$ to be able to keep enough states in the TDMRG
algorithm (with $M=100$ states). In the inset of Fig.~\ref{FigChi} we
show the convergence of the TDMRG method for several choices of the
Trotter time steps $\Delta \tau$ and numbers of states kept
for the example of $n_h=0.125$. Keeping only $M=45$ states is clearly
not sufficient, and this curve deviates from the other curves already
at rather high temperatures around $T\approx 0.3t$. The other curves
with $M=60$ and different values of $\Delta \tau$ show good
convergence, but at the lowest temperatures an accurate estimate of
the errors gets more difficult. In the figure for all hole densities
we therefore only show $\chi$ down to temperatures where we estimate
the truncation errors to be smaller than $1\%$ and interpolation
errors to constant particle density also smaller than $1\%$.
\figchi
The first observation is that both the symmetric and asymmetric model
show a strong enhancement of $\chi$ at temperatures near the gap value
of the undoped chain $T\propto J/2$. This strong enhancement of $\chi$
is caused by the creation of a small FM cloud around each mobile hole
by the double exchange mechanism, as the holes can gain kinetic energy
by FM alignment of the neighboring spins. At even lower temperatures,
the magnetic susceptibility is suppressed for the symmetric model due
to the spin gap and we find $\chi \rightarrow 0$ for $T\rightarrow
0$. In contrast, $\chi$ is finite and rather large at $T=0$ for the
asymmetric model, indicating the formation of larger FM moments due to
the double exchange mechanism (see ref.~\citen{Spin1}). However,
the ground state is spin singlet for both models, as we have tested
with the DMRG method by the calculation of $\sigma(i,j)=S_i^z
S_j^z-1/2(S_i^+ S_j^-)$ which vanishes for a rotationally invariant
ground state and the result is zero within the numerical precision of
the DMRG for this quantity for both models ($|\sigma(i,j)|<10^{-6}
\forall i,j$).

\subsection{Spin correlation function}
Complementary information about the magnetic properties can be gained
from the spin-spin correlations $S_{\text{sp}}(i-j)=\langle S^z_iS^z_j
\rangle$. For gapful systems we expect the spin-spin correlations to
decay exponentially, whereas we expect power-law decay for gapless
systems. By the DMRG method we have measured the spin-spin
correlations for the different models on very large chains of up to
$2\times 256$ sites, and we show some of those calculations in
Fig.~\ref{FigSpSp}. For the undoped system, the spin-spin correlations
clearly exhibit the expected exponential decay, and by fitting to the
asymptotic form $S_{\text{sp}}(l)\propto (-1)^l l^{-1/2} e^{-l/\xi}$
we obtain $\xi=6.05(5)$ in agreement with the results for the Haldane
$S=1$ chain \cite{White}.
\figsp
For both gapful symmetric models we expect exponentially decaying
spin-spin correlations also in the doped case. However, in
Fig.~\ref{FigSpSp} we observe an additional spiral-order in
$S_i^zS_j^z$ on top of the exponential decay upon doping, and we fit
to $S_{\text{sp}}(l)\propto \cos(2k_F l) e^{-l/\xi}$ in that case. The
correlation lengths listed in Table~\ref{TabGap} demonstrate that
correlation lengths $\xi \approx6.1(2)$ for $n_h=0.0625$ and $\xi
\approx 6.7(2)$ for $n_h=0.0625$ are very close to the undoped case in
the absence of the Coulomb repulsion, while the correlation length
$\xi$ increases by a factor of $2.56$ to $\xi \approx 15.4(5)$ for
$n_h=0.0625$ and $3.41$ to $\xi \approx 20.5(5)$ for $n_h=0.125$ at
$U=10t$. Rather surprisingly, also for the gapless asymmetric model
the spin-spin correlations seem to decay exponentially with
correlation lengths $\xi=7.9(2)$ for $n_h=0.0625$ and $\xi=11.2(2)$
for $n_h=0.125$. This exponential decay of the spin-spin correlations
at intermediate distances reflects the underlying spin-liquid
background, and we expect a crossover to power-law decay at larger
distances. However, since this length scale is given by the
polaron-polaron distance, we can not discriminate between power-law
and exponential decay due to the fact that even our largest systems
contain few polarons only, and much larger systems would be required
for that purpose.

This hierarchy of energy scales in the spin-sector of the asymmetric
model should also be visible by two peaks in the specific heat $c_V$.
First, we expect a large, broad peak around the gap-value $\Delta_s$
of the undoped $S=1$ chain stemming from the spin-liquid background,
and a second peak at low temperatures reflecting the lower gapless AF
interactions among the polarons. This two-peak structure is evident in
Fig.~4 of ref.~\citen{Spin1} by a broad peak around $T\approx J$
reflecting the spin-liquid background and from the steep increase of
$c_V$ signaled at the lowest temperatures. We estimate an energy scale
of $T<0.02t$ for the magnetic interactions among the polarons.
%

%
\subsection{String correlation function}
The hidden $Z_2\times Z_2$ symmetry of the Haldane $S=1$ chain is
revealed by the string correlation function $g(l)=\langle (S_{x_0}^z)
\left( \prod_{k=x_0+1}^{x_0+l-1} e^{i \pi S_{k}^z} \right)
(S_{x_0+l}^z)\rangle $, which quickly approaches the value
$g(l)\approx -0.374$ for $l\gg 1$ in the undoped case \cite{StringOrder}.
As a natural extension for the doped models we consider
\BEQ{EqString}
	g(x)=\left\langle ( \sum_{i=1,2} S_{x_0,i}^z) 
	\left( \prod_{k=x_0+1,
		j=1,2}^{x-1} e^{i \pi S_{k,j}^z} \right) 
	(\sum_{l=1,2} S_{x,l}^z )\right\rangle ,
\end{equation}
and the results are shown in Fig.~\ref{FigString}.
\figstring
In the undoped case we quickly recover the original result of
$g(l)\approx -0.374$ for $l\gg 1$. The value of $g(l)$ is only
slightly reduced to $g(l)\approx -0.31(2)$ for $n_h=0.0625$, and
$g(l)\approx -0.26(2)$ for $n_h=0.125$ for the symmetric model without
Coulomb repulsion in agreement with the weak-coupling result of
ref.~\citen{Fujimoto}. Combined with the very small deviation
in the spin correlation length $\xi$ from the undoped case, we conlude
that the spin structure of the symmetric model is very similar to that
of the undoped $S=1$ Haldane chain. The suppression of the string
correlation function $g(l)$ is much stronger if we include the Coulomb
repulsion or for the asymmetric model and we obtain $g(l)<0.1$ for
both of these models for $n_h=0.125$. In addition, the latter two
models also exhibit large oscillations in the amplitude of $g(x)$, caused
by the Friedel oscillations in the charge density due to the open
boundary conditions. By comparison $g(l)$ is constant within $1.2\%$
for the symmetric model without Coulomb repulsion apart from boundary
effects. In fact, for the asymmetric model we again expect a
crossover to power-law decay of $g(l)$ after sufficiently many
oscillations.

\subsection{Friedel oscillations and correlation exponent}
Friedel oscillations are induced in the charge density due to the open
boundary conditions. For gapful Luther-Emery and gapless
Tomonaga-Luttinger liquids, we can then obtain from these oscillations
the correlation exponent $K_\rho$ which determines the dominant
correlation functions in the thermodynamic limit.

We first consider the Fourier-transform of the charge density shown in
the inset of Fig.~\ref{FigCh}. For both of the symmetric models we
find only one peak at $k=2n\pi$. Since the lower band is fully
occupied for low hole densities, there are $N^{\text{up}}_h=2(1-n)L$
holes in the upper band and its particle density is
$n^{\text{up}}=2n-1$. For large Fermi volume, the Fermi vector is thus
given by $2k_F=\pi+(2n-1)\pi = 2n\pi$ and the above peak is compatible
with $2k_F$ oscillations.
\figch
For the asymmetric model we find only one peak at $k=n_h \pi$ at the
smallest hole density $n_h=0.0625$, but two peaks at $k=n_h \pi$ and
$k=2n_h \pi$ for $n_h=0.125$ as is evident in Fig.~\ref{FigCh}. Since
we denote the hole density in the conduction band with $n_h$ in this
case, we obtain a Fermi vector of $2k_F=\pi+(1-n_h)\pi \equiv n_h \pi$
for a large Fermi volume and the above peaks are compatible with
$2k_F$ and $4k_F$ oscillations.

After having identified the $2k_F$ and $4k_F$ fluctuations, we
determine the correlation exponent by fitting to the Friedel
oscillations of an impurity potential. In the gapful case they are
given by $\delta_n(x)\propto C_1 \cos(2k_F x) x^{-K_\rho} + C_2
\cos(4k_F x) x^{-2K_\rho}$ and $\delta_n(x)\propto C_1 \cos(2k_F x)
x^{-(1+K_\rho)/2} + C_2 \cos(4k_F x) x^{-2K_\rho}$ in the gapless case
\cite{Fabrizio}. For both gapful and gapless universality classes, a
value of $K_\rho>1$ leads to dominant pairing correlations in the
thermodynamic limit, while the CDW correlations decay slowest for
$K_\rho<1$. From Fig.~\ref{FigCh} it is evident that the amplitudes of
the symmetric model without Coulomb repulsion decay much faster than
for the other models where they are almost constant.  For the fit we
have to discard the states next to the boundary because of trapped
states. For the symmetric model without Coulomb repulsion we find by
fitting to the gapful Luther-Emery liquid $K_\rho\approx 1.5
\pm 0.05$ for $n_h=0.125$, showing dominant pairing correlations, and
$K_\rho \approx 0.41 \pm 0.08$ upon the inclusion of a Coulomb
repulsion of $U=10t$, giving dominant CDW correlations. For the
asymmetric model we fit to the gapless Tomonaga-Luttinger liquid and
find $K_\rho\approx 0.51 \pm 0.05$, also at $n_h=0.125$. Pair
formation can be explained by the gain of the large Hund's rule
coupling for holes pairs formed on the rungs of the symmetric model.
This energy gain is not possible for both of the other models, and in
contrast the holes may gain kinetic energy by the double exchange
mechanism in these cases. In the low hole doping region we can
determine hole pairing from the pair binding energy obtained from
$\Delta_{\text{pair}}=2 E_1-E_0-E_2$, with $E_n$ denoting the
ground-state energy with $n$ holes. For the symmetric model without
Coulomb repulsion we confirm hole pairing by a large binding energy of
$\Delta_{\text{pair}}\approx 2.29t$. Including the Coulomb repulsion,
the holes repel each other and we obtain $\Delta_{\text{pair}}\approx
-0.07t$, while the asymmetric model shows a very small positive
binding energy of $\Delta_{\text{pair}}\approx0.016t$ near
half-filling. Pair binding has also been found in ref.~\cite{Riera} in
a similar model.

\subsection{Pairing correlations}
Independently of the Friedel oscillations in the charge density, we
can also determine the correlation exponent from the pairing
correlations defined as $P_{i}(\vec{f'})P_{i+x}^\dagger(\vec{f})$,
where $P_{i}^\dagger(\vec{f})=\frac{1}
{\sqrt{2}}(c_{i,\uparrow}^\dagger c_{i+\vec{f},\downarrow}^\dagger \mp
c_{i,\downarrow}^\dagger c_{i+\vec{f},\uparrow}^\dagger)$ either
denotes the singlet ($-$) or the triplet pair ($+$) creation
operator. The pairing correlations also give further information on
the form-factor $\vec{f}$.  For gapful models belonging to the
universality class of Luther-Emery liquids, the pairing correlations
decay as $P_{i}(f)P_{i+x}^{\dagger}(f)\propto x^{-1/K_\rho}$
\cite{LutherEmery}, whereas for gapless Tomonaga-Luttinger liquids the
asymptotic form is given by $P_{i}(f)P_{i+x}^\dagger(f) \propto
A_0\ln(x)^{-1.5}x^{-1-1/K_\rho} +A_2\cos(2k_Fx)x^{-K_\rho-1/K_\rho}$
\cite{TomonagaLuttinger}.

\figpair
In Fig.~\ref{FigPair} we show the singlet pairing correlations
obtained from our DMRG calculations on systems with $2\times 128$
sites.  From the numerous possibilities of the form-factor $\vec{f}$
for the symmetric model, we restrict the discussion to those with the
largest amplitudes in the following.  In excellent agreement with the
previous results for $K_\rho$ we find $K_\rho\approx 1.55 \pm 0.05$
for the symmetric model without Coulomb repulsion at $n_h=0.125$. The
form factor of the pairs with largest amplitudes $\vec{f}=(1,1),
(2,1)$ and $(3,1)$ are consistent with a $d_{x-y}$ symmetry analogue
for a ladder. Surprisingly there are also almost no $2k_F$
oscillations in the pairing-correlations for the largest amplitudes,
despite the Friedel oscillations in the charge density.



Upon the inclusion of the Coulomb repulsion, the pairing correlations
decay much faster and in agreement with the previous result from the
Friedel oscillations we obtain $K_\rho\approx 0.42$. Finally by
simultaneously fitting to $f=2,4,6$, and $8$ for a Tomonaga-Luttinger
liquid, we obtain $K_\rho\approx 0.51 \pm 0.05$ for the asymmetric
model at $n_h=0.125$. The form factor of the pairing correlations with
the largest amplitudes is $f=8$ and $f=6$ in this case. Rather
extended pair formation is also evident from a corresponding, weak
substructure at a distance of $d=8$ in the hole pockets of the charge
density distribution in Fig.~\ref{FigCh}.

\figtriplpair
The triplet pairing correlations decay exponentially for both
universality classes. The correlations for the form factors $f$ with
the largest amplitudes are shown in Fig.~\ref{FigTripl} for systems
with $2\times 128$ sites and $P_{i}^\dagger(\vec{f})=\frac{1}
{\sqrt{2}}(c_{i,\uparrow}^\dagger c_{i+\vec{f},\downarrow}^\dagger +
c_{i,\downarrow}^\dagger c_{i+\vec{f},\uparrow}^\dagger)$. The
correlation functions for the other triplet states are similar. At
very short distances $d<5$, triplet pairing correlations for rung
pairs with $\vec{f}=(0,1)$ and $\vec{f}=(1,1)$ have the largest
amplitudes for the symmetric model without Coulomb repulsions, but
since they decay exponentially fast, the amplitudes for singlet pair
correlations become larger for distances $d>5$. Including the Coulomb
repulsion $U=10t$, the singlet correlations always have larger
amplitudes. For the asymmetric model, the triplet correlations have
the largest amplitudes for short distances $d\lesssim 5$ with a
form-factor of $f=4$, but again, the singlet correlations become
dominant at larger distances $d>5$ due the exponential decay of the
triplet correlations.

\subsection{Ferromagnetism and triplet pairing}
Ferromagnetism induced by the double-exchange mechanism is to be
expected for the asymmetric model and the symmetric model with Coulomb
repulsion at $J=J_d=0$, and might extend to small, but finite values of
$J,J_d$. Rather large FM regions have been identified in the
phase diagrams of refs.~\citen{Riera,Kubo} for related models.

In a first approach we determine the FM region by comparing
ground-state energies in different total $S^z_{\text{tot}}$ subspaces
calculated by the DMRG method. In the FM phase, all the spins are
fully aligned and the ground state is compatible with
$S^z_{\text{tot}}=(2L-N_h)/2$, while for singlet ground state with
$N_h$ even, $S^z_{\text{tot}}=0$ is lowest in energy, where $N_h$
denotes the number of holes in the system of length $L$. Because of
the number of subspaces with different $S^z_{\text{tot}}$ becomes very
large for large system sizes, we can do such calculations only for
rather small systems, and we have used $L=32$ for our
calculations. For the asymmetric and the symmetric model with $U=5t$,
$S^z_{\text{tot}}=0$ is lowest in energy for $J=J_d\gtrsim 0.1t$, and
$S^z_{\text{tot}}=(2L-N_h)/2$ is lowest for $J=J_d=0$. For
intermediate values $0<J,J_d<0.1t$ some
$0<S^z_{\text{tot}}<(2L-N_h)/2$ is lowest in energy. While these
results clearly demonstrate a singlet groundstate for $J=J_d \gtrsim
0.1t$ and a FM groundstate for $J=J_d=0$, the results are not
conclusive for the intermediate region, due to the special boundary
conditions required to suppress the chain-end excitations.

An alternative approach to determine the FM region is possible by the
spin-spin correlations. Working in the $S^z_{\text{tot}}=0$ subspace,
the spins are ferromagnetically aligned in the $x-y$ plane in the FM
phase.
\figspsm
In Fig.~\ref{FigSpSm} we show the $S^+(2)S^-(x)$ correlations for the
asymmetric model and the symmetric model with $U=5t$ calculated by the
DMRG method for a system size of $L=32$ sites. In the asymmetric
model, all the spins are fully ferromagnetically aligned for
$J=J_d=0$. Ferromagnetic alignment of the spins is still evident for
$J=J_d=0.01t$, however at the end of the chain, the last spin is
antiferromagnetically aligned. For $J=J_d=0.1t$, we find AF alignment
compatible with the singlet groundstate. The same situation is
encountered also for the symmetric model with $U=5t$, where we find
fully polarized spins for $J=J_d=0.01t$, a few spins at the boundary
with AF aligment for $J=J_d=0.05t$, and AF alignment for $J=J_d=0.1t$,
where the period of the AF alignment corresponds to $2k_F$
oscillations. Without Coulomb repulsion, the FM phase seems to be
slightly reduced, since the strong pair binding on the rungs reduces
the double exchange mechanism. Already for $J=J_d=0.01t$ a few spins
are antiferromagnetically aligned in this case, but the bulk is still
ferromagnetically aligned. The AF alignment of the spins at the
boundaries is expected to be a finite-size effect only and can be
explained by the decreased mobility of the holes near the boundaries,
which in turn reduces the double-exchange mechanism and favors AF
order. At $J_H=10t$ we therefore expect a narrow FM region for all
models for $0\leq J=J_d \lesssim 0.1t$.

\figFMPair
The possiblity of triplet superconductivity in the FM phase is a
question of great interest and we have investigated this problem by
the pairing correlation shown in Fig.~\ref{FigFMPair}. As is evident
from the figure, the triplet pairing correlations for the symmetric
model with $J_H=-10t$, and $J=J_d=0.01t$ at $n_h=0.125$ for a triplet
pair $P_{i}^\dagger(\vec{f})=\frac{1}
{\sqrt{2}}(c_{i,\uparrow}^\dagger c_{i+\vec{f},\downarrow}^\dagger +
c_{i,\downarrow}^\dagger c_{i+\vec{f},\uparrow}^\dagger)$ formed on a
rung are roughly three orders of magnitude larger than the largest
singlet pairing pairing corrletions, and at least for this system
size, do not seem to decay faster than the singlet pairing
correlations. Also for the asymmetric with the same parameters, the
triplet pairing correlations are roughly three orders of magnitude
larger than the singlet pairing correlations, and do not seem to decay
faster than the singlet pairing correlations. From the system sizes in
our study it is not possible to distinguish between exponential or
power-law decay, but at least for the $2\times32$ systems, dominant
triplet pairing seems possible. Larger system sizes are needed in
order to determine which correlation function decays slowest in the
thermodynamic limit of infinite system size.

\section{Conclusions}
In our study we have investigated three realistic models of AF $S=1$
Heisenberg chains doped with mobile $S=1/2$ fermions in the strong
coupling regime: a model with a level difference between the two
orbitals forming the $S=1$ spins, a model without level difference,
and a model without level difference including a strong Coulomb
repulsion between the electrons in the two orbitals on the same
site. Our investigation by numerical methods shows very different
physical properties of these models.

For the asymmetric model with a level difference between the two
orbitals, the spin gap is destroyed immediately upon doping, while it
remains finite for both symmetric models with equal particle density
in both orbitals. This collapse of the spin gap in the asymmetric
model is caused by AF interactions among the polarons. The polarons
are created by the double-exchange mechanism in order to gain kinetic
energy, and if the AF couplings between the neighboring sites are
$J,J_d=0$, the ground state is FM. We also find evidence for a FM
ground state for very small values of $J,J_d\lesssim 0.05t$ for all
models. The absence of interactions among the polarons in the
symmetric model lead to a finite spin gap in that case.

However, even for the asymmetric model the holes are only a weak
perturbation of the underlying spin liquid. A hierarchy of energy
scales is thus given in the spin sector of the asymmetric model by the
lowest-lying, gapless AF interactions among the polarons and second,
larger energy scale of the order of the spin gap of the undoped system
by the underlying spin liquid which remains intact. As a consequence
of the spin liquid background, the spin-spin correlations decay
exponentially fast for all models for the system sizes in our study
(up to $2\times 256$ sites). However, we expect a crossover to
power-law decay for the asymmetric at large distances. Also the string
correlation function, which quickly approaches a nonzero value of the
long-ranged order in the undoped system, remains finite for all
models. Again, we expect a crossover to power-law decay for the
asymmetric model at large distances.

Since hole pairing on the rungs is only possible for the symmetric
model without strong Coulomb repulsion, we find completely different
dominant correlations functions in thermodynamic limit of infinite
system size. The symmetric model shows dominant pairing correlations
and $K_\rho>1$ without Coulomb repulsion, in contrast to dominant CDW
correlations with $K_\rho<1$ for the symmetric model with Coulomb
repulsion and for the asymmetric model. In the FM region for small
values of the AF interactions $J,J_d\lesssim 0.05t$, the triplet
pairing correlations are strongly enhanced.

\section*{Acknowledgements}
We wish to thank H. Asakawa and H. Tsunetsugu for valuable
discussions. The numerical calculations have been performed on
workstations at the ISSP. This work is supported by the ``Research for
the Future Program'' (JSPS-RFTF 97P01103) from the Japan Society
for the Promotion of Science (JSPS).


%


\begin{thebibliography}{999}
\bibitem{Haldane} F.D. Haldane: Phys. Lett. {\bf 93A} (1983) 464;
	 \PRL{50} (1983) 1153.
\bibitem{StringOrder} M. den Nijs and K. Rommelse: \PRB{40} (1989) 4709;
	S.M. Girvin and D.P. Arovas: Phys. Scr. {\bf 27} (1989) 156;
	H. Tasaki: \PRL{66} (1991) 798.
\bibitem{DblExch} C. Zener: Phys. Rev. {\bf 82} (1951) 403; 
	P. W. Anderson and H. Hasegawa: Phys. Rev. {\bf 100} (1955) 675;
	P.-G. de Gennes: Phys. Rev. {\bf 118} (1960) 141.
\bibitem{DiTusa} J.F. DiTusa {\it et al.}: \PRL{73} 1857 (1994).
\bibitem{YBaNiO} D.J. Buttrey, J.D. Sullivan, and A.B. Rheingold:
	J. Solid State Chem. {\bf 88} (1990) 291; J. Amador {\it et
	al.}: \PRB{42} (1990) 7918; R. S\'{a}ez-Puche {\it et al.}:
	J. Solid State Chem. {\bf 93} (1991) 461; J. Darriet and
	L.P. Regnault: Solid State Commun. {\bf 86} (1993) 409;
	B. Battlogg, S.-W. Cheong, and L.W. Rupp, Jr.: Physica B {\bf
	194-196} (1994) 173.
\bibitem{Kojima} K. Kojima {\it et al.}: \PRL{74} (1995) 3471.
\bibitem{Sorenson} E.S. S\mbox{\o}renson and I. Affleck: 
	\PRB{51} (1995) 16115.
\bibitem{Kaburagi} M. Kaburagi, I. Harada, and T. Tonegawa: 
	J. Phys. Soc. Jpn. {\bf 62} (1993) 1848; 
	M. Kaburagi, and T. Tonegawa: J. Phys. Soc. Jpn. {\bf 63} (1993) 420.
\bibitem{Kennedy} T. Kennedy: J. Phys. Condens. Matter {\bf 2} (1990) 5737.
\bibitem{Penc} K. Penc, and H. Shiba: \PRB{52} (1995) R715.
\bibitem{Dagotto} E. Dagotto, J. Riera, A. Sandvik, and A. Moreo:
	\PRL{76} (1996) 1731; C.D. Batista, A.A. Aligia, and J. Eroles:
	\PRL{81} (1998) 4027; 
	E. Dagotto, and J. Riera: \PRL{81} (1998) 4082.
\bibitem{Riera} J. Riera, K. Hallberg, and E. Dagotto: Phys. Rev. Lett. 
	{\bf 79}, (1997) 713.
\bibitem{Spin1} B. Ammon and M. Imada: Phys. Rev. Lett. (to be published).
\bibitem{Frahm} H. Frahm, M. P. Pfannm\"{u}ller, and A. Tsvelik:
		\PRL{81} (1998) 2116;
		H. Frahm and S. Sobiella: \PRL{83} (1999) 5579;
		H. Frahm and N. A. Slavanov: preprint cond-mat/9912319.
\bibitem{Fujimoto} S. Fujimoto and N. Kawakami: \PRB{52} (1995) 6189.
\bibitem{Nagaosa} N. Nagaosa and M. Oshikawa: J. Phys. Soc. Jpn {\bf 65}
	 (1996) 2241.
\bibitem{LevelDiff} B. Ammon and M. Imada: J. Phys. Soc. Jpn {\bf 69}
		(to be published).
\bibitem{TDMRG} R.J. Bursill, T. Xiang, and G.A. Gehring:
		J. Phys. Cond. Matt. {\bf 8} (1996) L583;
		X. Wang and T. Xiang: \PRB{56} (1997) 3177; 
		N. Shibata: J. Phys. Soc. Jpn. {\bf 66} (1997) 2221.
\bibitem{White} S.R. White: \PRL{69} (1992) 2863.
\bibitem{WhiteFullDMRG}	S.R. White: \PRB{48} (1993) 10345.
\bibitem{TM}  H. Betsuyaku: Prog. Theor.
		Phys. {\bf 73} (1985) 319; T. Yokota and H. Betsuyaku:
		Prog. Theor. Phys. {\bf 75} (1986) 46.
\bibitem{trotter_suzuki} H.F. Trotter: 
              Proc. Am. Math. Soc. {\bf 10} (1959) 545; 
              M.\ Suzuki: Prog. Theor. Phys. {\bf 56} (1976) 1454.
\bibitem{troyer_tm} M. Troyer, H. Tsunetsugu, and D. W\"urtz: 
	Phys. Rev. B {\bf 50} (1994) 13515.
\bibitem{BiOTDMRG} B. Ammon, M. Troyer, T.M. Rice, and N. Shibata:
	\PRL{82} (1999) 3855.
\bibitem{Fabrizio} M. Fabrizio and A.O. Gogolin: \PRB{51} (1995) 17827;
	R. Egger and H. Grabert: \PRL{75} (1995) 3505;
	R. Egger and H. Schoeller: Czech. J. Phys. {\bf 46} (1996) Suppl. S4,
	1909.
\bibitem{LutherEmery} A. Luther and V.J. Emery: Phys. Rev. 
        Lett. {\bf 33} (1974) 589.
\bibitem{TomonagaLuttinger} J. S\'olyom, Adv. Phys.: {\bf 28} (1979) 201;
	V.J. Emery: {\it Highly Conducting One-Dimensional Solids},
	 (Plenum, New York 1979), edited by J.T. Devreese {\it et al.}.
\bibitem{Kubo} K. Kubo {\it et al.}: preprint cond-mat/9811286.
\end{thebibliography}
\end{document}